\documentclass[preprint,12pt]{elsarticle}

\usepackage{amssymb}

\usepackage[version=3]{mhchem} 
\usepackage{siunitx} 
\usepackage{graphicx} 
\usepackage{natbib} 
\usepackage{amsmath} 
\usepackage{color, colortbl}
\usepackage{soul}
\usepackage{lscape}
\usepackage{tikz}
\def\checkmark{\tikz\fill[scale=0.4](0,.35) -- (.25,0) -- (1,.7) -- (.25,.15) -- cycle;}

\definecolor{Gray}{gray}{0.9}
\newcolumntype{L}[1]{>{\raggedright\let\newline\\\arraybackslash\hspace{0pt}}m{#1}}
\newcolumntype{C}[1]{>{\centering\let\newline\\\arraybackslash\hspace{0pt}}m{#1}}
\newcolumntype{R}[1]{>{\raggedleft\let\newline\\\arraybackslash\hspace{0pt}}m{#1}}

\journal{Computers and Security}

\begin{document}

\begin{frontmatter}



\title{Enhancing Enterprise Network Security: Comparing Machine-Level and Process-Level Analysis for Dynamic Malware Detection}


\author[inst1]{Baskoro Adi Pratomo}
\ead{baskoro@if.its.ac.id}

\affiliation[inst1]{organization={Informatics Department, Institut Teknologi Sepuluh Nopember},
            city={Surabaya},
            state={East Java},
            country={Indonesia}}

\author[inst2]{Toby Jackson}
\author[inst2]{Pete Burnap}
\author[inst2]{Andrew Hood}
\author[inst2]{Eirini Anthi}

\affiliation[inst2]{organization={School of Computer Science and Informatics, Cardiff University},
            city={Cardiff},
            state={Wales},
            country={United Kingdom}}

\begin{abstract}
Analysing malware is important to understand how malicious software works and to develop appropriate detection and prevention methods. Dynamic analysis can overcome evasion techniques commonly used to bypass static analysis and provide insights into malware runtime activities. Much research on dynamic analysis focused on investigating machine-level information (e.g., CPU, memory, network usage) to identify whether a machine is running malicious activities. A malicious machine does not necessarily mean all running processes on the machine are also malicious. If we can isolate the malicious process instead of isolating the whole machine, we could kill the malicious process, and the machine can keep doing its job. Another challenge dynamic malware detection research faces is that the samples are executed in one machine without any background applications running. It is unrealistic as a computer typically runs many benign (background) applications when a malware incident happens. Our experiment with machine-level data shows that the existence of background applications decreases previous state-of-the-art accuracy by about 20.12\% on average. We also proposed a process-level Recurrent Neural Network (RNN)-based detection model. Our proposed model performs better than the machine-level detection model; 0.049 increase in detection rate and a false-positive rate below 0.1.
\end{abstract}

\begin{keyword}
malware analysis \sep dynamic analysis \sep event log \sep sysmon \sep process-level data
\end{keyword}

\end{frontmatter}


\section{Introduction}
Malware attacks are prevalent nowadays. According to Statista~\cite{statista_2022}, there were 5.4 billion malware attacks in 2021 and 2.8 billion attacks in the first half of 2022. To tackle the problem of malware infection, researchers and security companies have developed various solutions to detect malicious activities. A common approach for detecting malware in a system is by analysing the file's content. For example, by calculating the hash value of a file and comparing it with a database of known malicious hashes; or by examining the codebase. 

The aforementioned approaches have two problems. Firstly, they assume that all malware can be located by examining a malicious file. There have been some incidents~\cite{crowdstrike_2022} that involve file-less malware. This kind of malware implants itself into a specific process in memory; thus, examining hash values and code bases is less effective for identifying known malware signatures. Secondly, both approaches rely on a database of known malicious hashes or malware signatures. They cannot detect malware that has never previously been seen.

Machine learning (ML) techniques have been proposed to help identify malware without depending on code or hash file analysis. ML works by analysing various malware and benignware features. For example, previous approaches have analysed machine-level resource usage (e.g., CPU, memory, disk usage) during malware and benignware execution. Then, those features are used to see whether the machine was running malware or not~\cite{ucci2019survey}, \cite{rhode2018early}. 

Machine-level data captures the usage of a machine while opening and running applications, files etc. Previous work has collected machine data in aggregate form - combining all the various applications and activities on a machine into one observation and determining if malicious activity is present on a second-by-second basis \cite{rhode2018early}. In reality, malware may be running along with other benignware on a machine simultaneously. It is unlikely that a machine would have malware running on its own without other benign processes running in parallel. Therefore, the machine-level data does not solely capture malicious behaviour but will also be affected by benign process activities.

In this paper we argue that process-level data captures the behaviour of each running process and as such, provides a more granular view of which  activities are malicious. We can separate the benign and malicious activities of a machine. Process-level data  also captures information such as what processes are created, which registry entries are modified, what files are created or modified, what domain names are contacted, and many others. 

In this research, we conduct a novel investigation into process-level features for malware detection and compare these results with machine-level features. We argue that the existence of background applications during malware executions will affect the performance of the existing model. Therefore, we proposed that process-level data would be more suitable for detecting malware activities.

Additionally, datasets used in previous research on dynamically detecting malware using ML used only a single virtual machine to generate their malicious and behavioural activity - frequently without any other background applications running. In an enterprise environment, this is unrealistic, casting doubts over the applicability of previously tested methods in practice. Typically, we would have multiple computers running various applications, some of which might be malicious. For this reason, in this research, we created a virtualised small-medium enterprise topology to emulate a real-world network, executed both malware and benign samples across multiple machines - alongside typical benign activity, and then captured both machine and process-level data from each machine on the network. The generated data there enabled us to evaluate the first approach to study process-level data for malware detection while also doing so in a much more realistic and diverse multi-endpoint virtual network.

In summary, the main contributions of this paper are as follows:
\begin{itemize}
    \item The first ML-based malware detection model to determine whether a specific process is malicious by using process-level data
    \item A malware dataset containing both machine-level and process-level data from benign and malicious samples. The data also include second-by-second information, making it possible to evaluate the behaviour of malware execution over time.
    \
\end{itemize}

The rest of the paper is structured as follows: Section \ref{sec:2-related-work} discusses related work in malware datasets and detection. We discuss our data generation methodology and detection model development in Section \ref{sec:3-methodology}. The results of our experiments are presented in Section \ref{section:4-result}. We present some issues and the limitations of our approach in Section \ref{section:5-issues}. Lastly, the paper concludes in Section \ref{section:6-conclusion}.

\section{Related Work}
\label{sec:2-related-work}
In the area of malware datasets, we already have several datasets with millions of samples, such as Ember~\cite{anderson2018ember}, Solem-20M~\cite{harang2020sorel20m}, and MOTIF~\cite{joyce2022motif}. The EMBER dataset was generated by extracting features from malicious and benign PE files using the LIEF project~\cite{LIEF}. SoRel-20M was produced by Sophos and contained disarmed malicious PE files. Besides the disarmed malicious PE files, SoRel-20M also provides features from the PE files, which were extracted using EMBER's \texttt{PEFeatureExtractor}. Similarly, the MOTIF dataset \cite{joyce2022motif} contains 3,095 disarmed malicious PE files and is labelled with malware family labels. It also provides EMBER's raw features from its samples. 

Despite having many samples, the aforementioned datasets only extracted their features from static analysis of the malicious and benign files. Despite its data richness, features obtained from the static analysis may be limited as the malware may hide its characteristics by using static analysis evasion techniques. On the other hand, the dynamic analysis may provide better insight into malware behaviour while running. None of the aforementioned datasets executed their samples and captured the sample behaviour.

\cite{rhode2018early} generated a malware and benignware dataset by executing both malicious and benign samples in a virtualised environment using Cuckoo~\cite{cuckoo_sandbox}. While the sample was executed, they captured machine-level data (i.e., CPU usage, memory usage, network usage, and the number of processes). They also developed a Recurrent Neural Network model that can predict malicious behaviour early within five seconds. 

Similar to \cite{rhode2018early}, \cite{sihwail2019malware} generated a malware dataset by executing samples with Cuckoo. This research, however, does not provide data on CPU, memory, and network usage, but the authors captured the API calls conducted by the analysis machine during execution. 

All malware datasets \cite{rhode2018early, sihwail2019malware} that were generated by dynamic analysis executed their samples in a single virtual machine or sandbox. This is to ensure malware execution is contained and does not spread beyond the analysis environment so that the analysis environment can quickly be reset to its original state without needing to reinstall the entire operating system. However, in reality, malware and benignware may interact with other machines, affecting the captured behaviour in other network parts. Therefore, in this research, we executed malware samples simultaneously as benignware being executed on different machines on the same network - creating a more realistic environment with more interaction between machines and more background noise. 

Table \ref{tab:comparison} shows the difference between our malware dataset and the previous research. None of the previous research added background noise (i.e., benignware running simultaneously on a different machine) and captured process-level data.

\begin{table}[]
\label{tab:comparison}
\caption{Comparison of existing malware datasets and our proposed dataset}
\begin{tabular}{l|C{0.2\textwidth}|C{0.2\textwidth}|C{0.2\textwidth}}
\hline
Dataset  & Malware Analysis & Background Applications & Process-level Data \\
\hline \hline
EMBER~\cite{anderson2018ember} &  Static                &                  &                    \\
Sorel20M\cite{harang2020sorel20m} &  Static                &                  &                    \\
MOTIF\cite{joyce2022motif} &  Static                &                  &              \\     
Rhode, et al\cite{rhode2018early} &  Dynamic                &                  &              \\
Sihwail, et al\cite{sihwail2019malware} & Dynamic                  &                  &              \\
Our dataset & Dynamic & \checkmark & \checkmark
\end{tabular}
\end{table}

Dynamic analysis has been one of the two methods to identify malware. Looking at the malware behaviour when running can provide us with more information than merely looking at the static code. Apart from generating a dataset, we also evaluated how machine-level and process-level data for dynamic analysis can benefit malware detection. This is because capturing malicious behaviour from the machine-level information, such as CPU, memory, and network usage - as per \cite{rhode2018early} - may not capture sufficient information to identify the difference between malicious behaviour and benign or background processes. For that reason, our dataset also obtained Sysmon events that were generated by all processes such that we can know exactly what each application on the network was doing during execution.

\cite{vinayakumar2019robust} and \cite{kumar2019malware} each proposed a malware detection method using API calls. They looked at the list of API calls made by the applications and identified malicious behaviour from the sequence of API calls. \cite{rosli2020ransomware} approached malware detection using the files created by the malware and developed a graph-based detection model accordingly. These approaches are based on process-level detection, but none look at the events created by the applications.

Other research analyses Windows events generated by a process to find malicious software running on the system~\cite{tobiyama2016malware}. However, their approach only captured file creation, registry value set, and thread creation. Moreover, they did not analyse the effect of running background applications during sample executions. In comparison, our proposed method analysed the effect of background applications and considered various events; hence ours is more comprehensive. Table \ref{tab:comparison-method} summarises the differences between our proposed method and the previous works.

\begin{table}[]
\label{tab:comparison-method}
\caption{Comparison of existing dynamic malware detection and our proposed method}
\begin{tabular}{l|C{0.2\textwidth}|C{0.4\textwidth}}
\hline
Research  & Background Applications & Input Features \\
\hline \hline
Rhode, et al\cite{rhode2018early} & - & CPU, memory, network usage \\ \hline
Vinayakumar, et al~\cite{vinayakumar2019robust} & - & API calls                    \\ \hline
Kumar, et al~\cite{kumar2019malware} & - & API calls                    \\ \hline
Tobiyama, et al~\cite{tobiyama2016malware} & -  & File creation, Thread started, and Registry value set \\  \hline
Our proposed method & \checkmark & Process creation, Network connection, File creation, Process termination, Registry value set and modified \\ \hline
\end{tabular}
\end{table}

\section{Methodology}
\label{sec:3-methodology}
The section that follows is divided into two main parts. The first part details how we obtained the dataset required for the research. We accomplished this by executing both malware and benign samples in a controlled environment to simulate a real-world scenario. We collected the data generated by these executions, which included Windows Events such as process creation, network connection, file access, and registry access, among others.

The second part of this section describes the development of the Recurrent Neural Network (RNN) model (i.e., Long Short-Term Memory~\cite{hochreiter1997lstm} and Gated Recurrent Unit~\cite{chung2014gru}) that we used to predict malicious processes early using their associated Windows Events as input features. We picked RNN-based approaches as they are suitable for handling a sequence, making them particularly useful for processing a sequence of Windows Events. We preprocessed the dataset to remove any noise or irrelevant data and then trained the RNN model using the remaining features. Our model was designed to learn from benign and malicious processes' patterns and classify unknown processes as benign or malicious based on their generated sequence of events.

\subsection{Data Generation}
\label{sec:data-generation}
We generated the data by running malicious and benign software in a virtualised environment to compare the machine-level and process-level information. Then, we captured the machine utilisation (i.e., CPU, memory, network usage, and the number of processes) and the Sysmon events while  samples were running. 

\begin{figure}
    \centering
    \includegraphics[width=\textwidth]{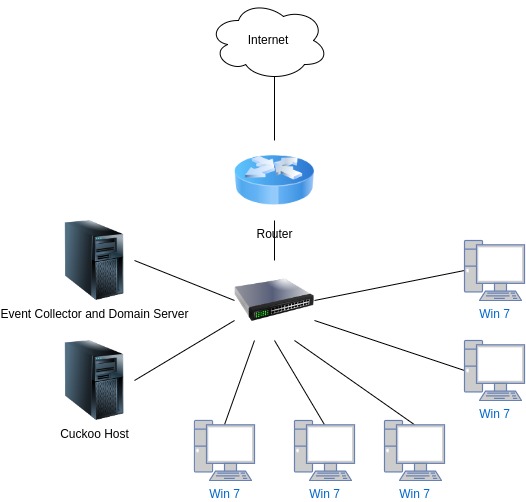}
    \caption{The network topology used in the experiment. All outgoing traffic is blocked. The network contains five Windows 7 SP1 machines}
    \label{fig:experiment_network}
\end{figure}

The data generation experiment was conducted in a simulated network environment to represent a typical small to medium enterprise computer network~\cite{jenkins2003secure}. The network consisted of five Windows 7 operating systems machines, a Windows Server 2016 machine that served as the Sysmon event collector, and a Linux machine that was used to detonate malware samples using Cuckoo Sandbox~\cite{cuckoo_sandbox}. All these machines are connected through a layer-2 switch, as shown in Figure \ref{fig:experiment_network}.

To ensure that our data was accurate and reliable, we carefully configured each machine in the network environment. We recorded the specific operating systems, IP addresses, and installed applications used in each machine in Table \ref{tab:client_subnets}. We also applied all necessary Windows updates to prevent Sysmon from generating duplicate Process GUIDs, which could have made it difficult to identify each process. The updates ensured that each process was uniquely identified by its GUID, and we were able to accurately track the events generated by each process during our experiments.

During the data generation phase, we executed both benign and malicious samples on the network environment and collected the Windows Events generated by each execution. The events collected included process creation, network connection, file access, and registry access, among others. This process allowed us to obtain a large and diverse dataset, which we used to develop and train our Recurrent Neural Network (RNN) model. The RNN model was specifically designed to analyze the patterns of Windows Events generated by both benign and malicious processes and to accurately classify unknown processes based on their event sequences. Overall, the data generation experiment was carefully designed to ensure that our dataset was representative of real-world scenarios and that our analysis was based on accurate and reliable data.

\begin{table}[]
    \centering
    \caption{Required OS and applications in each server}
    \begin{tabular}{L{0.2\textwidth}|L{0.2\textwidth}|L{0.3\textwidth}|L{0.2\textwidth}}
        \textbf{Host} & \textbf{OS} & \textbf{Installed applications} & \textbf{IP addresses} \\ \hline
        \rowcolor{Gray}
        Client Subnet & Windows 7 with KB3033929 and KB4457144 updates for Sysmon Process Guid bug fixes & Web browsers (Chrome), Acrobat Reader 9, Hollows Hunter, Sysmon, Python2.7 with psutil and pillow & 172.16.5.(101-105) \\
    \end{tabular}
    \label{tab:client_subnets}
\end{table}

\begin{figure}
    \centering
    \includegraphics[width=\textwidth]{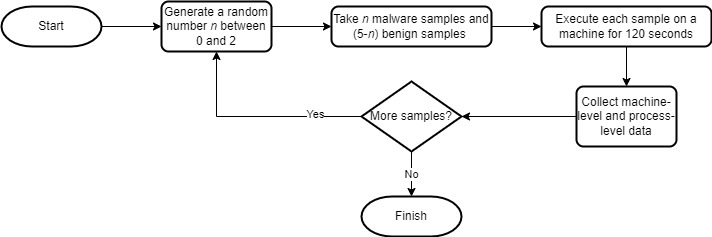}
    \caption{The flowchart of the data generation methodology}
    \label{fig:flowchart}
\end{figure}

Logging Made Easy (LME) is an open-source initiative that combines various freely available software components to offer foundational security information logging on Windows devices. It simplifies the process of integrating with a Security Information and Event Management (SIEM) system. Thus, it serves as our data collection system for process-level information as it enables efficient logging and monitoring of security-related data which is based on Sysmon~\cite{russinovich_sysmon} and Windows Event Forwarder~\cite{windowseventforwarder}.

The LME Event Collector server had several responsibilities, including managing the analysis machines through Group Policy and handling DNS requests. We applied three primary policies to all analysis machines to enable effective data collection and malware analysis. First, we elevated a domain user to a local administrator to facilitate certain administrative tasks. Second, we enabled Windows Event Log forwarding to ensure that the generated Windows Events were collected in a centralized location. Finally, we disabled both Windows Firewall and Defender to prevent malware from being blocked or deleted before it could be executed.

These policies were critical to the success of our experiment, as they ensured that the Windows Events were accurately collected and stored in one location. Disabling Windows Firewall and Defender enabled the malware to run without interference, allowing us to analyze its behaviour and generate the necessary data for our research. Overall, the LME Event Collector server played a crucial role in managing the analysis machines and facilitating our data collection and analysis processes.

LME Event Collector collects Windows events from the analysis machines every fifteen seconds. The period of fifteen seconds was chosen to ensure all events would be collected before the VM reset to a clean state. We collected Sysmon event ID 1 (Process Creation), 2 (File Creation Time Changed), 3 (Network Connection), 5 (Process Termination), 7 (Image Load), 8 (Remote Thread Creation), 11 (File Creation), 12, 13, 14 (Registry-related events), and 22 (DNS queries). The server's ability to collect events was provided by Logging Made Easy~(LME)~\footnote{available at https://github.com/ukncsc/lme}, which utilises Windows Event Forwarding.

We executed malware by using Cuckoo to send files or binaries to a client machine. All logs from Cuckoo were then sent to the Cuckoo Server to be analysed later. As both the Cuckoo server and the client machines are running on virtual machines, we configured Cuckoo to use \emph{physical} machine settings. We also modified Cuckoo to shut down VMs after it finished executing applications since the cyber range on which the virtualised network runs resets the VM state only when it is shut down. The shutting down process was actioned by calling a stop API request to the cyber range platform and restarted by making an API request to the cyber range platform. This activity would be recorded in the network traffic, but as of now, we did not capture any network traffic.

Apart from its default behaviour log. We also set Cuckoo to collect the machine utilisation (i.e., CPU, memory, network usage, and the number of processes) with the script from \cite{rhode2018early} and to run two instances of Hollows Hunter. Hollows Hunter is used to scan for process hollowing; a technique commonly used to hide malicious processes. By default, it scans all active processes in the machine, but as some malware may run for a few milliseconds and thus evade detection, we ran two instances of Hollow Hunter, one for scanning all processes and the other for scanning the injected samples. Hollows Hunter log files were then sent to Cuckoo Server at the end of each sample execution.

\begin{table}[]
    \centering
    \caption{Required OS and applications on machines in the Logging and Management subnet}
    \begin{tabular}{L{0.2\textwidth}|L{0.1\textwidth}|L{0.3\textwidth}|L{0.3\textwidth}}
        \textbf{Server} & \textbf{OS} & \textbf{Installed applications/services} & \textbf{IP address/netmask} \\ \hline
        \rowcolor{Gray}
        Cuckoo Server & Ubuntu 20.04 & Cuckoo  & 172.16.5.50/24 \\ 
        LME Event Collector & Windows Server 2016 & Logging Made Easy (LME) & 172.16.5.10/24 \\
    \end{tabular}
    \label{tab:management}
\end{table}

Cuckoo Sandbox operates by receiving a list of filenames that are to be injected into virtual machines. The Cuckoo daemon schedules the order in which the files are injected into each virtual machine. If multiple files are being injected into different virtual machines, their tasks can be executed simultaneously. In our experiment, we had five virtual machines available, which allowed us to execute up to five samples (both benign and malicious) simultaneously. We referred to each round of execution as an \emph{iteration}. For instance, if we had a total of twenty samples, we would complete four \emph{iterations} to execute all the samples.

Before picking which malware and how many of them would be running at the same time, for each \emph{iteration}, we picked a random number between zero to two (approx half of the number of machines). The next \emph{iterations} started by picking a new random number from zero to two again. In the case of getting zero for three times in a row, the next \emph{iteration} should pick any number greater than zero and less than two.

Each \emph{iteration} in the experiment involves running five samples with the random number of malware samples executed in that \emph{iteration}. For this research, only malicious binaries are picked. We also tried to make the malicious samples to be diverse by including various type of malware. However, a malicious sample can belong to more than one category. Therefore, the number of each malware type is not balanced. Each sample was then executed into a randomly chosen VM. 

In each set of experiments, we initially had 200 malware and 200 benignware. In total, there are 1195 samples, a similar number of samples to the previous works~\cite{rhode2018early, sihwail2019malware}. The malicious samples are obtained from VirusShare~\cite{virusshare.com}, while the benign samples came from the previous research~\cite{rhode2018early}. As the number of malware and benignware was the same and the number of malware executed in each iteration was not always two, some benign samples may be executed multiple times, but malicious samples were executed only once. We also ensured that each experiment had different malicious and benign samples. The exact number of unique malicious and benign samples for each experiment is shown in Table \ref{tab:samples} and the number of each malware variant is listed in Table \ref{tab:malware-variant}.

\begin{table}[htbp]
    \centering
    \caption{The number of malicious and benign samples for each experiment}
    \begin{tabular}{|l|l|l|r|}
        \hline
        \multicolumn{1}{|c|}{\textbf{Experiment}} & \multicolumn{1}{c|}{\textbf{Class}} & \multicolumn{1}{c|}{\textbf{Filetype}} & \multicolumn{1}{c|}{\textbf{\# of samples}} \\ \hline
        \multicolumn{ 1}{|c|}{Set-0} & \multicolumn{ 1}{c|}{Benign} & PE32 & 195 \\ \cline{ 2- 4}
        \multicolumn{ 1}{|l|}{} & \multicolumn{ 1}{c|}{Malicious} & PE32 & 200 \\ \hline
        \multicolumn{ 1}{|c|}{Set-1} & \multicolumn{ 1}{c|}{Benign} & PE32 & 190 \\ \cline{ 2- 4}
        \multicolumn{ 1}{|l|}{} & \multicolumn{ 1}{c|}{Malicious} & PE32 & 200 \\ \hline
        \multicolumn{ 1}{|c|}{Set-2} & \multicolumn{ 1}{c|}{Benign} & PE32 & 210 \\ \cline{ 2- 4}
        \multicolumn{ 1}{|l|}{} & \multicolumn{ 1}{c|}{Malicious} & PE32 & 200 \\ \hline
    \end{tabular}
\label{tab:samples}
\end{table}

\begin{table}[htbp]
\centering
\caption{The distribution of malware variants. The variant is not mutually exclusive. Some malware may belong to more than one category.}
\begin{tabular}{|l|r|r|r|}
\hline
\textbf{Malware Type} & \multicolumn{1}{l|}{\textbf{Set-0}} & \multicolumn{1}{l|}{\textbf{Set-1}} & \multicolumn{1}{l|}{\textbf{Set-2}} \\ \hline
Ransomware & 10 & 10 & 13 \\ \hline
Trojan & 198 & 200 & 199 \\ \hline
Botnet & 26 & 27 & 20 \\ \hline
Exploit & 2 & 0 & 5 \\ \hline
Miner & 18 & 23 & 21 \\ \hline
\end{tabular}
\label{tab:malware-variant}
\end{table}

Once each sample finished being analysed, we reset the VM, and Cuckoo would schedule the next sample to analyse immediately. Ideally, all samples finished at the same time, so the next five samples would start at the same time. However, there were some issues (explained in Section \ref{section:5-issues}) with the Cyber Range which caused some samples to be left behind and executed later. To solve this problem, we set the start\_time option in Cuckoo which allowed us to set the time before the analysis machine started executing the sample. Samples from the same iteration would always start at the same time in the record such that the malware samples will always be running at the same time as benignware which we refer to as \emph{background noise}. As we randomised the VMs where the malware was executed, it will also give the ML model a challenge as the model would not be able to identify malware based on merely the information where the malware was executed.

Each client machine was installed with standard office applications, such as browser, Word, Excel, Teams, Outlook, and PDF reader. We executed a sample for 120 seconds. After the time ran out, we sent Hollow Hunter log files and the machine utilisation records to Cuckoo Server before shutting down and resetting the machine. LME logs were collected at the end of each experiment by manually collecting them from the event collector server. Figure \ref{fig:flowchart} summarises the data generation process.

\begin{figure}
    \centering
    \includegraphics[width=\textwidth]{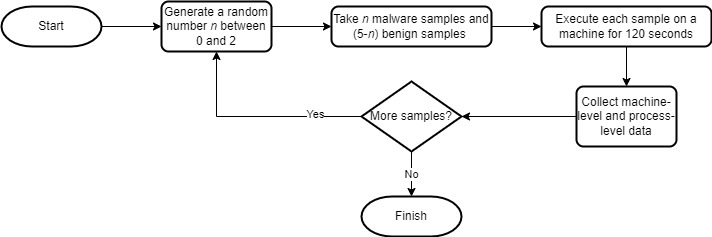}
    \caption{The flowchart of the data generation methodology}
    \label{fig:flowchart}
\end{figure}

\subsection{Malware Detection Model}
After collecting the data, we developed a detection model and re-evaluated the machine-level RNN model developed by \cite{rhode2018early}. Rhode et al.'s \cite{rhode2018early}'s RNN model reads system utilisation data every second and tries to predict whether a machine is running malware each second. The result shows that the detection gets more accurate over time. 

Our detection model builds on previous research by focusing on process-level data. Instead of detecting whether a particular machine is running malware in general, the model looks for specific malicious processes based on the sequence of events generated by a process and the data collected using Hollows Hunter~\cite{hasherezade_hollows_2022}.

As explained in Section \ref{sec:data-generation}, we did three experiments to generate malware activity data. We refer to the data generated by these experiments as Set-0, Set-1, and Set-2, respectively. Each set contains malicious and benign data, which will be used in subsequent experiments to evaluate the malware detection approaches. 

\subsection{Machine-level detection model}
\begin{figure}
    \centering
    \includegraphics[width=\textwidth]{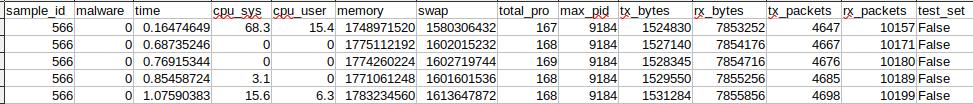}
    \caption{A part of system utilisation data from an execution of a benign sample}
    \label{fig:machine-level-data}
\end{figure}

The malware detection using system utilisation data is heavily based on the previous work by \cite{rhode2018early}. The previous work developed an RNN model that analysed system utilisation every second and gave an accurate early prediction after analysing the data for five seconds. The model observes the machine's CPU (system and user) usage, memory usage, swap usage, the total number of processes, maximum process ID, and the number of bytes and packets transmitted and received. Figure \ref{fig:machine-level-data} shows a part of system utilisation data from an execution of a benign sample. The system utilisation data were taken by Cuckoo every second, as was done in \cite{rhode2018early}. Also, note that we developed a single model to be used in all machines. There is no machine-specific model as all machines will share the same model. This is an additional enhancement to previous research. 

There is no difference in terms of the methodology, but it is worth noting that \cite{rhode2018early} executed malicious applications in their analysis machine without other analysis machines executing benign samples, while our dataset added \emph{background noise} during the analysis. Therefore, we expected to see performance degradation in the result as it should be harder for the model to distinguish between malware and benignware.

For this experiment, we combined Set-0, Set-1, and Set-2 and then split the data into training and testing sets. The RNN model is trained with 10-fold cross-validation on the training set. Then we evaluated the model by using the testing set. Hence, no training data is mixed with the testing data. We also followed \cite{rhode2018early}'s approach to measure the model's quality by using accuracy metrics and added precision, recall, and F1-score metrics for better comparison with the process-level data. 

\subsection{Process-level detection mode (LME Events and Hollows Hunter)}
Machine-level data can only tell us which machine is performing malicious activities. While this information might help identify the infected machine, it would be more beneficial to identify the specific malicious process. Knowing this enables us to shut down the specific process instead of the whole machine. In this research, we experimented with process-level data gathered from LME data and Hollows Hunter logs. The LME data contains important events which are generated by the process, while the Hollows Hunter data have the information on whether a particular process has potentially malicious implants (i.e., replaced/implanted PE files, shellcodes, hooks, or in-memory patches).

\subsubsection{Data Preprocessing}

\begin{figure}
    \centering
    \includegraphics[width=\textwidth]{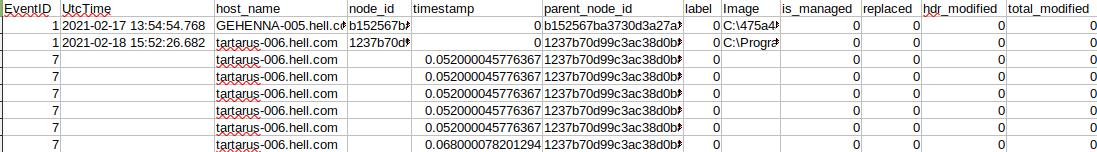}
    \caption{An example of a sequence of events generated by a benign process. Each event is depicted in a row. Some fields are empty because the event does not have that information. The events were taken directly from the .evtx files. Therefore, some unused attributes are still shown and have not been preprocessed.}
    \label{fig:process-level-data}
\end{figure}

LME data are essentially Windows Events stored in .evtx files. For easier handling, we converted the .evtx files to newline JSON format with evtx\_dump tool~\cite{evtx_dump}. The resulting JSON file contains a list of unordered events. For this research, we only considered event ID 1 (process creation), 3 (network connection), 5 (process termination), 12, and 13 (both are registry events). We then correlated the events by using ProcessGuid to look for a sequence of events generated by a process. As a result, we ended up with an event tree containing a list of process creation events and what the process did as shown in Figure \ref{fig:process-level-data}.

Although not all, some processes were identified containing implants by HollowsHunter, either malicious or not. The data needed to be incorporated with the LME data. As these data come from different sources, we identified the relationship between the HollowsHunter and the LME data by the process id, machine name, and timestamp.

Each type of event has a different set of attributes, but some of them are shared. To model these events into a vector with uniform features, we flattened all possible attributes of an event and filtered out unnecessary attributes. If an event does not have the attribute, such as a process creation event that does not have information about the network endpoint it is connected to, we filled the attribute for that particular event as N/A. We removed attributes that have too many distinct values or contain the filename as it might hint too much to the model that the vector is malicious or benign. For non-binary categorical attributes, we transformed the features with one-hot encoding. And the timestamp was transformed to the number of milliseconds after the initial process creation event.

Lastly, as a process may only have a set of numbers from Hollows Hunter, while it may generate multiple events, we repeated the Hollows Hunter data across the series of vectors of the particular process. In the end, we have 31 features, including the Hollows Hunter data and the timestamp. Table \ref{tab:process-level-features} shows the list of the features used in our process-level detection model. We also categorised our features based on the data source, i.e., LME and Hollows Hunter features.

\begin{table}[]
\centering
\caption{The features used in our process-level detection model. Grey cells are features from Hollow Hunters, and the rest is from LME data}
\begin{tabular}{|l|l|l|}
\hline
\rowcolor{Gray}
is\_managed & replaced & hdr\_modified  \\ \hline
\rowcolor{Gray}
total\_modified & patched & iat\_hooked  \\ \hline
\rowcolor{Gray}
implanted\_shc & unreachable\_file & other  \\ \hline
\rowcolor{Gray}
implanted\_pe & & \\ \hline
SameImageLoaded & DPortName\_https       & EventID\_3             \\ \hline
SignatureStatus & DPortName\_other       & EventID\_5             \\ \hline
Signed          & IntegrityLevel\_High   & EventID\_12            \\ \hline
Signed\_Failed  & IntegrityLevel\_Low    & EventID\_13            \\ \hline
Protocol\_udp   & IntegrityLevel\_Medium & EventType\_DeleteValue \\ \hline
Protocol\_tcp   & IntegrityLevel\_System & EventType\_SetValue    \\ \hline
DPortName\_http & EventID\_1             &                        \\ \hline
\end{tabular}
\label{tab:process-level-features}
\end{table}

We conducted the process-level detection model experiments using two different sets of features. The first set of features includes a sequence of events generated by a process. The second set of features contains more detailed information, such as the features listed in Table \ref{tab:process-level-features}. For the sake of brevity, we will refer to the first set of features as the Event-only feature set and the second set of features as the Complete feature set.

\subsubsection{Recurrent Neural Network process-level malware detection}

To identify malicious processes, we have implemented two types of recurrent neural network (RNN) models - a Long Short-Term Memory (LSTM) based model and a Gated Recurrent Unit (GRU) model. The reason behind using RNN models is that the data we have for our model can be represented as time series data and consists of varying lengths of events. Our primary objective is to create a baseline model that can be used for further research in this field. Using RNN models, we can capture the sequential patterns present in the data and accurately identify the malicious processes. We hope this model will help improve the accuracy and efficiency of identifying malicious processes.

Our RNN-based model takes input in the form of a sequence of events, denoted as $X = {x_i | 0 \leq i \leq n}$. The sequence length is represented by $n$, and each event $x_i$ is a vector that captures information about the event. In the Event-only features, we used one-hot encoding to represent each event as a vector with five elements, as our research only considered five types of events. On the other hand, in the Complete feature set, each event is represented as a vector with 32 elements, as listed in Table 6. This means that for each process, we have a time series of event vectors with one vector for each second. This allows our RNN-based model to capture the temporal dependencies between events and accurately identify malicious processes.

After selecting the appropriate set of features, we proceeded to build a one-layer Recurrent Neural Network-based model for classifying the sequence of events. In this model, each time step of the recurrent layer takes the event features as input. The RNN model captures the temporal dependencies in the sequence of events and outputs a hidden state at the final time step. This hidden state is then passed through a linear transformation layer with a Sigmoid activation function. During the training phase, the model's parameters are adjusted using backpropagation, which optimizes the model's ability to classify the input data accurately. During the identification/testing phase, the model is used to identify whether a given process is malicious or not. If the model's output exceeds 0.5, the process is classified as malicious. 

\section{Detection Model Performance}
\label{section:4-result}
This section discusses the result of the machine-level and process-level detection models explained in Section \ref{sec:3-methodology}. We evaluated both approaches on a second-by-second basis; the model performance is measured every second such that we know when the models start making good decisions. All experiments were run on a PC with Core i7 10700 2.9 GHz, 32 GB of RAM, NVIDIA GeForce RTX 2060, NVIDIA CUDA 10.0, and CUDNN 8.

During the data generation process, each machine generated a time-series system utilisation data. If the injected application was malware, the generated data were marked as malicious. By the end of the data generation process, we have a collection of time-series data for each machine and each application execution. We refer to the time-series data generated by a machine as \emph{machine activity} and the sequence of events generated by a process as \emph{process activity} - the latter being the novel element of the experimentation. 

True Positive (TP) represents the number of correctly classified malicious activities, and True Negative (TN) represents the number of correctly classified benign activities. At the same time, False Positive (FP) and False Negative (FN) represent the number of wrongly classified benign and malicious activities, respectively. We then measure the 
performance by using accuracy, precision, recall, and F1-score which are calculated as in Equation \ref{eq:acc}, \ref{eq:precision}, \ref{eq:recall}, and \ref{eq:f1} respectively.

\begin{equation}
    Acc  = \frac{\texttt{TP}+\texttt{TN}}{\texttt{TP}+\texttt{TN}+\texttt{FP}+\texttt{FN}} * 100
    \label{eq:acc}
\end{equation}

\begin{equation}
    Precision  = \frac{\texttt{TP}}{\texttt{TP}+\texttt{FP}}
    \label{eq:precision}
\end{equation}

\begin{equation}
    Recall  = \frac{\texttt{TP}}{\texttt{TP}+\texttt{FN}}
    \label{eq:recall}
\end{equation}

\begin{equation}
    F_1  = 2*\frac{Precision * Recall}{Precision+Recall}
    \label{eq:f1}
\end{equation}

We compared the machine-level detection model performance by running \cite{rhode2018early}'s model with our dataset which contains \emph{background noise} as one of the most recent and best-performing model to detect such malware is the one developed by Rhode et al. The best practice for evaluating a machine learning approach is to have separate training and testing set. As the name implies, the training set is used to train the model, and the testing set is for evaluating the model's performance. In this experiment, we combined Set-0, Set-1, and Set-2 then split them into the training and testing sets with a ratio of 80:20.

\begin{figure}
    \centering
    \includegraphics[width=\textwidth]{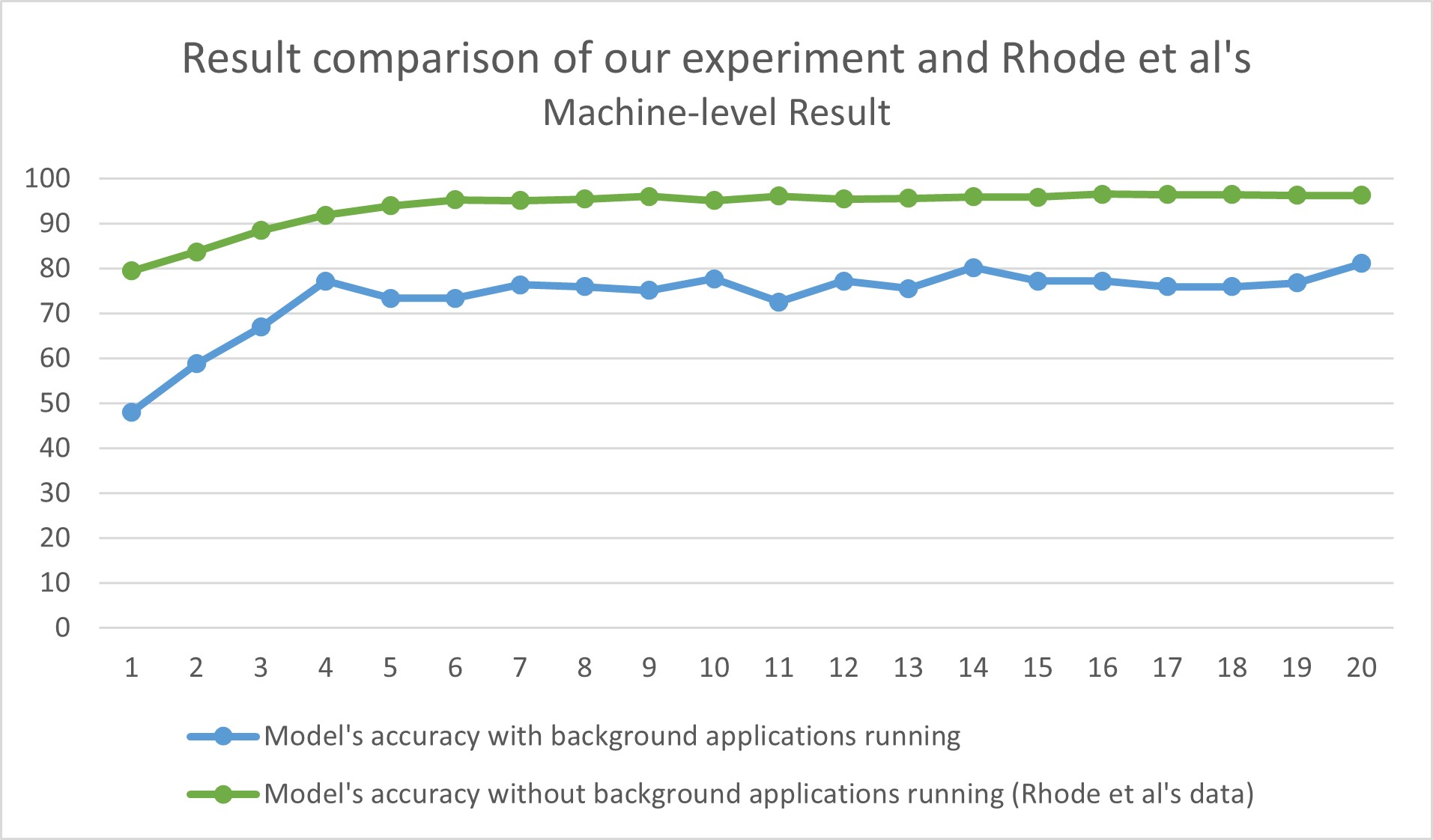}
    \caption{Result comparison of the machine-level activity RNN model with the new data that contain background applications and the data from \cite{rhode2018early} that do not.}
    \label{fig:machine-level-result}
\end{figure}

Table \ref{tab:machine-act-result} shows the result of the detection of both models for the first twenty seconds, with the last column being the result taken from the per-second result in \cite{rhode2018early}. As shown in Table \ref{tab:machine-act-result} and Figure \ref{fig:machine-level-result}, \cite{rhode2018early}'s result tends to be more accurate over time, particularly during the first five seconds. The results from our new experiment (the remaining columns) show that the model shows similar behaviour to \cite{rhode2018early} during the first five seconds. The accuracy sees an increase and then becomes relatively plateaued. The accuracy result is also confirmed by the other metrics (i.e., precision, recall, and F1-score) showing the same trend. However, our accuracy is always below \cite{rhode2018early}'s result. We argue that this decrease in performance is caused by the existence of benign applications running at the same time as the malware. The inclusion of additional benign samples injected into the virtual environment at the same time malware samples were executed is one key difference between our data set and \cite{rhode2018early}'s. The addition of multiple processes running in parallel, while more representative of real-world systems, clearly impacts the performance of the RNN approach - presenting a new research challenge of distinguishing between malicious and benign activity - where previous research tended to only inject malware for dynamic analysis - with no background noise. 

\begin{table}[!ht]
    \centering
    \caption{The results of the RNN model for machine-level data. Only accuracy is presented as the previous work only have that metric. It is shown that having background applications running decreases the model performance}
    \begin{tabular}{|L{0.1\textwidth}|C{0.4\textwidth}|C{0.4\textwidth}|}
    \hline
        \textbf{Time step} & \textbf{Accuracy with background applications (\%)} & \textbf{Accuracy without background applications (\%)} \\ \hline
        1 & 48.07 & 79.5 \\ \hline
        2 & 58.80 & 83.69 \\ \hline
        3 & 66.95 & 88.48 \\ \hline
        4 & 77.25 & 91.92 \\ \hline
        5 & 73.39 & 93.95 \\ \hline
        6 & 73.39 & 95.28 \\ \hline
        7 & 76.39 & 95.12 \\ \hline
        8 & 75.97 & 95.48 \\ \hline
        9 & 75.11 & 96.02 \\ \hline
        10 & 77.68 & 95.11 \\ \hline
        11 & 72.53 & 96.13 \\ \hline
        12 & 77.25 & 95.46 \\ \hline
        13 & 75.54 & 95.6  \\ \hline
        14 & 80.26 & 95.93 \\ \hline
        15 & 77.25 & 95.87 \\ \hline
        16 & 77.25 & 96.54 \\ \hline
        17 & 75.97 & 96.5 \\ \hline
        18 & 75.97 & 96.43  \\ \hline
        19 & 76.82 & 96.26 \\ \hline
        20 & 81.12 & 96.26 \\ \hline
    \end{tabular}
    \label{tab:machine-act-result}
\end{table}


After experimenting with the machine-level data, we continue with the process-level data. We developed our detection model with LSTM and GRU. Both models used ADAM as the optimiser with a learning rate of 0.01. We set the loss function to binary cross entropy. The models were developed with Python 3.8.10 and PyTorch 0.2.0 library. 

Our dataset contains an imbalanced ratio of benign and malicious process events, with a greater number of benign events than malicious ones. To ensure balance between the two classes in our training and testing sets, we performed undersampling on the benign class. We randomly split the malicious samples with a ratio of 80:20. Then, we took the same number of benign samples for the training set and used the rest of the benign samples for the testing set. We did that because the proportion of benign and malicious samples is imbalanced. In summary, we have 420 malicious and benign samples for the training set and 105 malicious and benign samples for the testing set.


We repeated all experiments ten times with randomly picked samples for the training and testing sets and averaged the results. Unlike the machine activity-level model, we pay more attention to the precision, recall, and F1 score as we would like to get more insight from the result. Precision measures the ratio of the correctly malicious detected samples to the number of samples being classified as malicious, as formulated in equation \ref{eq:precision}. Recall measures the number of malicious samples correctly detected as malicious, as formulated in equation \ref{eq:recall}. Some literature refers to recall as the detection rate. F1-score conveys the balance between precision and recall, as formulated in equation \ref{eq:f1}.

\begin{figure}
    \centering
    \includegraphics[width=\textwidth]{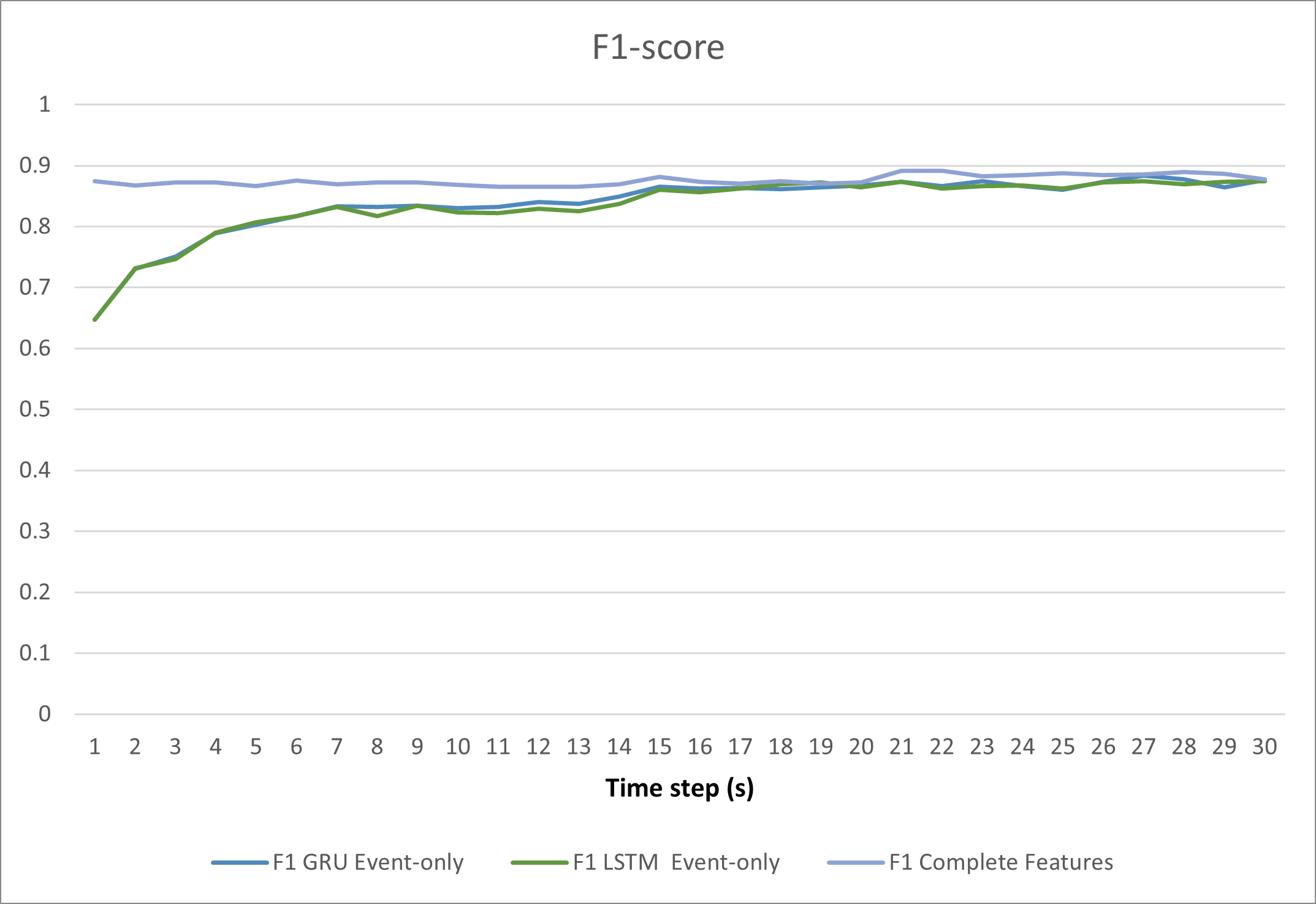}
    \caption{The F1-score of our proposed model when it was trained and evaluated with the Event-only and the Complete feature set}
    \label{fig:f1-process-result}
\end{figure}

\begin{figure}
    \centering
    \includegraphics[width=\textwidth]{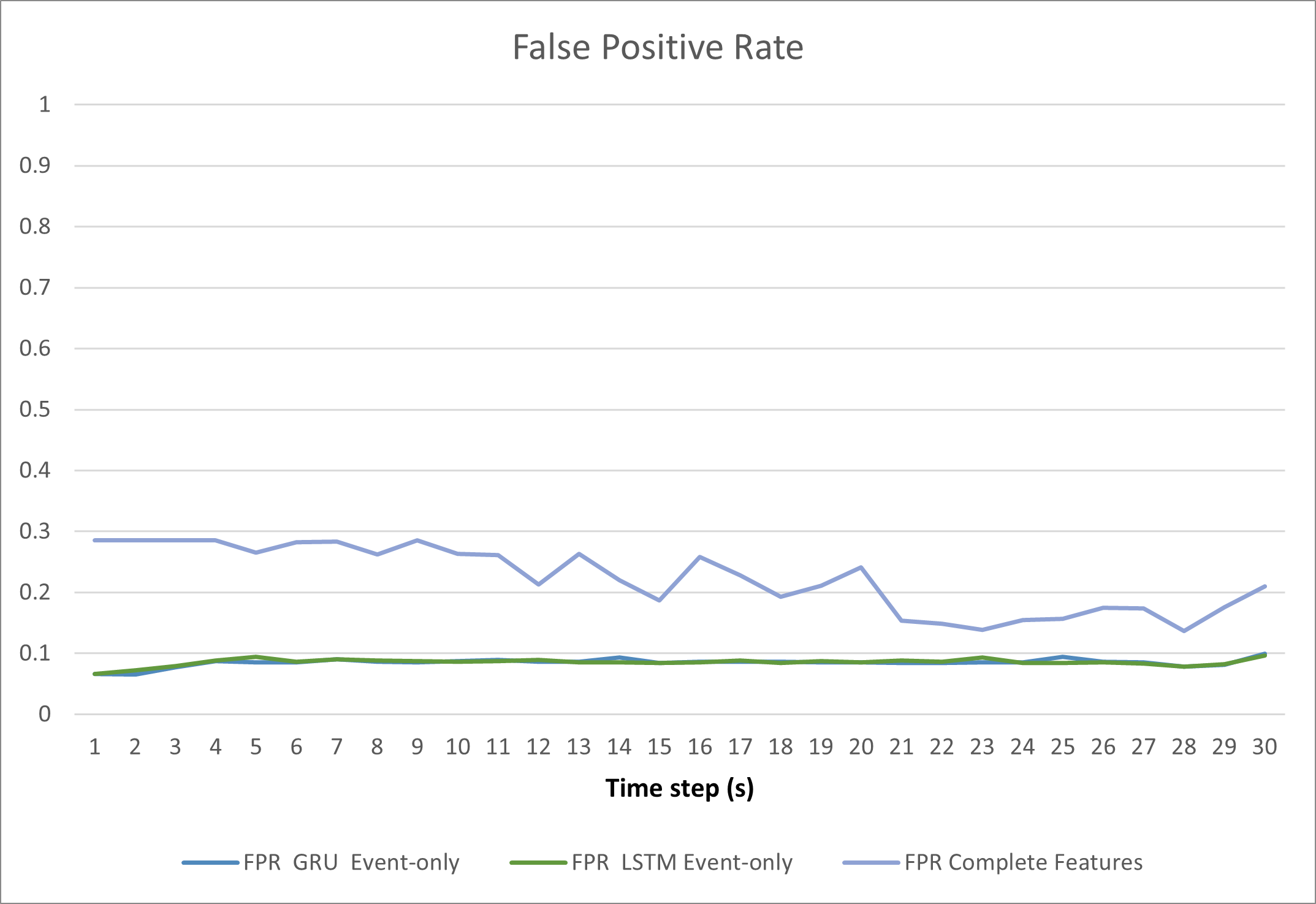}
    \caption{The false positive rate (FPR) of our proposed model when it was trained and evaluated with the Event-only and the Complete feature set}
    \label{fig:fpr-process-result}
\end{figure}

\begin{figure}
    \centering
    \includegraphics[width=\textwidth]{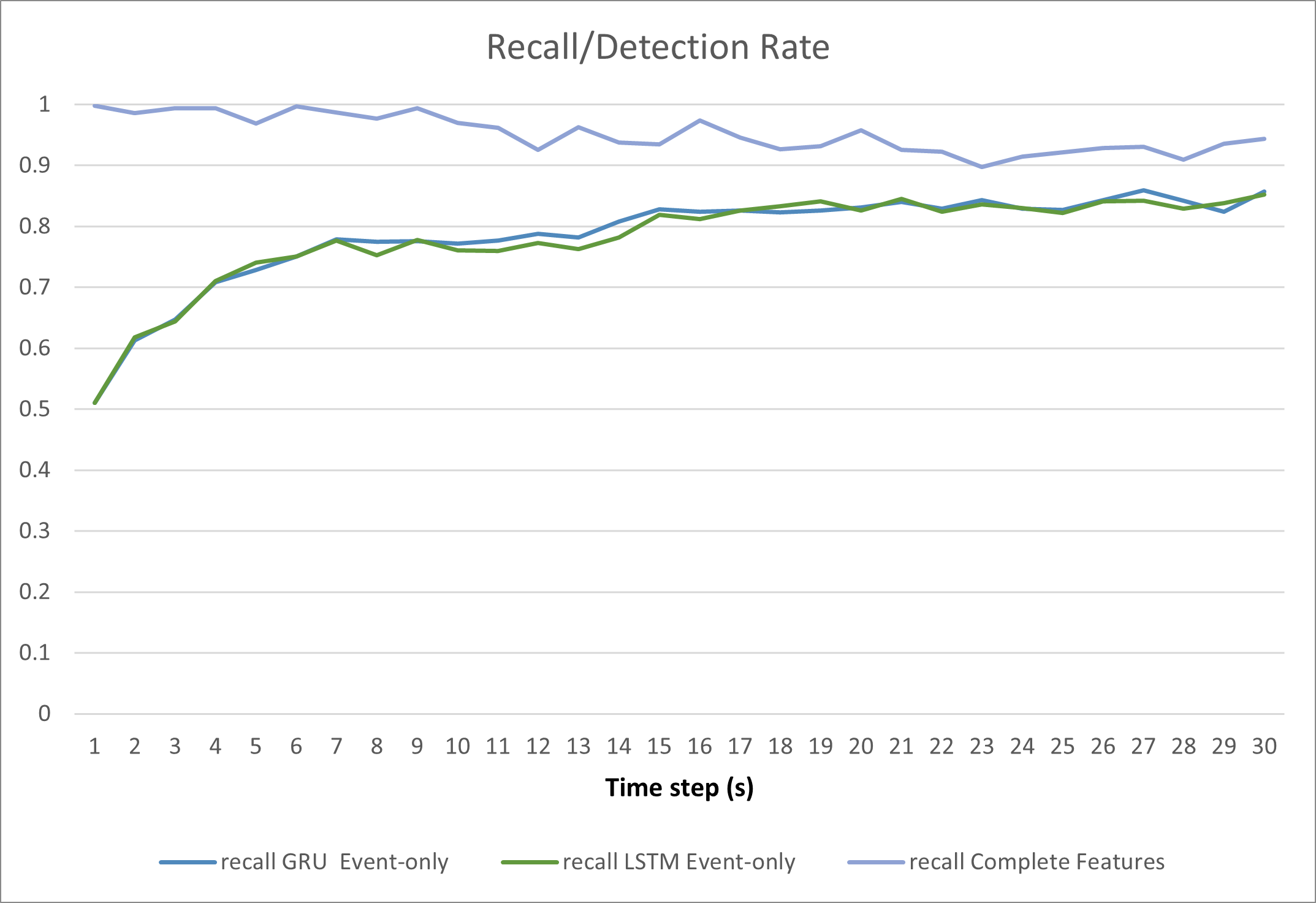}
    \caption{The recall/detection rate of our proposed model when it was trained and evaluated with balanced and imbalanced sets}
    \label{fig:recall-process-result}
\end{figure}

As shown in Figure \ref{fig:f1-process-result}, the Complete feature set provides stable performance with an F1 score of 0.87, while the Event-only feature set starts with 0.65 and keeps increasing over time. The performance of the Event-only feature set stops increasing after fifteen seconds. However, the Event-only feature set has a lower false positive rate (FPR) (see Figure \ref{fig:fpr-process-result} than the Complete Feature set. It stays below 0.1, while the FPR of the Complete set is greater than 0.2 despite the value decreasing over time. Figure \ref{fig:recall-process-result} also shows that the detection rate of the Complete feature set decreases over time. It is interesting because typically performance will increase when we have more data coming in.



\begin{figure}
    \centering
    \includegraphics[width=\textwidth]{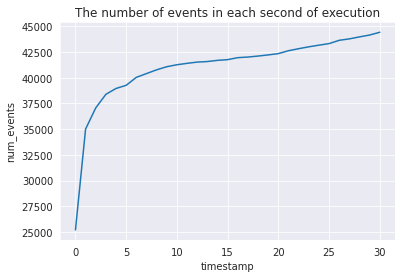}
    \caption{The number of events from the initial sample execution time to thirty seconds afterwards.}
    \label{fig:num-events}
\end{figure}

\begin{figure}
    \centering
    \includegraphics[width=\textwidth]{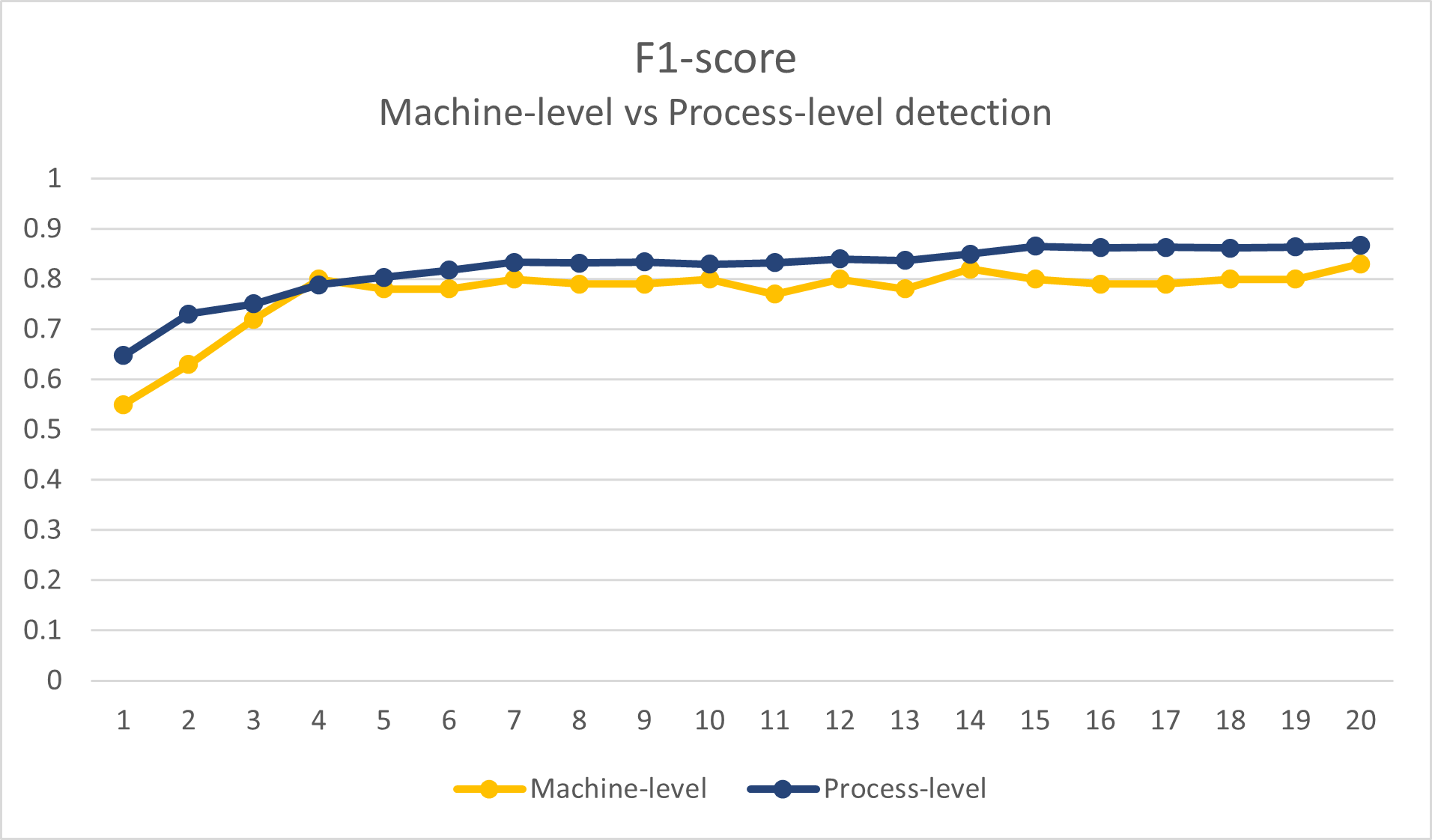}
    \caption{The comparison of machine-level and process-level detection. The F1-score shows that our process-level detection model performed better than the machine-level one since the first second.}
    \label{fig:f1-machine-process}
\end{figure}

Should we compare the performance of the machine-level and the process-level with the Event-only feature set detection model (see Figure \ref{fig:f1-machine-process}). We can see that the process-level detection model performs better than the machine-level one from the first second (see Figure \ref{fig:num-events}). The machine-level performance never surpasses the process-level despite more data coming in.

Another point worth noting is that the effect of Hollows Hunter features on the model's performance. In our dataset, only 101 out of 1200 samples have features extracted from Hollows Hunter. The F1-score between the model which considered Hollows Hunter features and the model which did not consider them only differs by 0.01 on average. Therefore, the added Hollows Hunter features do not seem to significantly affect the model's performance.

And as can also be seen in Figure \ref{fig:f1-process-result}, using either LSTM or GRU does not give a significant impact on the performance. Both RNN-based models always provide similar F1-score.


\section{Issues and Limitations}
\label{section:5-issues}
We faced several issues in our two experiments that need to be taken into account for future works, particularly if our work is going to be reproduced. We executed the data generation process on Cardiff University's Cyber Range which is based on Hynesim~\cite{hynesim} and Qemu~\cite{bellard2005qemu}. Most of these issues are related to the behaviour of the Qemu.

All client machines were configured to be immutable, which means that all changes to the machine will be removed when it is shut down. Restarting the machine keeps the changes. The problem is Cuckoo restarts guest machines after each analysis. Therefore, we modified Cuckoo to shut down guest machines and turn them back on by sending an API request to the Cyber Range platform. From our experiments, the second step was not always successful. The shutdown process is a non-blocking process. There was a time when Cuckoo sent the API request before the machine entirely shut down, which caused the request to be ignored. To handle that problem, we then ran another script alongside the Cuckoo daemon to monitor the machine's state. The script sends another API request to turn on inactive machines.

However, that script does not fully solve the issue as another issue arose when some virtual machines had been turned off and on many times. The Cyber Range failed to start some machines, and sometime later, all client machines failed to start, including other machines in the Cyber Range. When this happens, the only possible solution is to reset the platform, but it will stop and undefine all virtual machines. This issue might cause disturbance when many people are using the cyber range. Although the Cuckoo daemon can automatically continue from where it left off, as mentioned earlier, some samples were executed later than their counterparts.


Another issue we faced during the data collection was duplicate ProcessGuid in the data we obtained from the LME. ProcessGuid is supposed to be a unique value that can be used to correlate events. According to Sysmon documentation~\cite{russinovich_sysmon}, the value is generated by combining machine GUID, process ID, and timestamp. However, we found out the root causes of this problem. The problem was caused by missing Windows updates (KB3033929 and KB4457144) which made Sysmon improperly generate zeros in the middle part of the ProcessGuid. 


\section{Conclusion and Future Work}
\label{section:6-conclusion}
This research has generated a malware activity dataset containing machine-level data (i.e., system utilisation) and process-level (i.e., LME and Hollows Hunter data). The data was generated in a small enterprise network-like environment to understand better how malware propagates across networks, as none of the previous research has considered it. 

We also experimented with detection models that are trained on machine-level and process-level data. The result from the machine-level detection model shows a performance drop (on average 20.12\% in accuracy) compared to earlier work~\cite{rhode2018early}. It shows that background applications may affect detection performance. 

Our RNN-based model with the process-level data provides better performance than the machine-level data; 0.049 average increase in detection rate and  false-positive rate below 0.1. The detection performance keeps increasing significantly until we have seven seconds of process-level activities. The performance grows slower shortly afterwards. However, better feature engineering is still needed for future research in process-level malware detection.

We can pursue several other directions as a follow-up of this research. We executed our malware samples by sending a sample to the analysis machine and waiting for 120 seconds. We only assumed the adversary would merely drop the malware into the victim's machine. In reality, the story might be more complex as there are usually several infiltration steps. The adversary may also make a lateral movement after the initial malware infection. This behaviour is not captured in our dataset. To have this kind of behaviour, we suggest using Mitre Caldera~\cite{mitrecaldera} to emulate adversarial activities.

Another thing that could be improved is the way we run background applications. In our setup, we ran the background applications after the user logged in, and then there was no user interaction. We let the background applications stay idle. It would be more realistic to emulate user behaviour, e.g., typing in a Word document, browsing the internet, opening a PDF file and interacting with it. 

 \bibliographystyle{elsarticle-num} 
 \bibliography{references}





\end{document}